\documentclass[prb,showpacs,preprintnumbers,amsmath,amssymb,twocolumn]{revtex4}

\usepackage{graphicx}
\usepackage{dcolumn}
\usepackage{bm}

\begin{document}

\preprint{APS/123-QED}

\title{Long range diffusion noise in platinum  micro-wires with metallic adhesion layers}

\author{Z. Moktadir$^*$}
\affiliation{School of Electronics and Computer Science,
Southampton University, UK.}
\author{ J. W. van Honschoten}
\affiliation{Transducers Science and Technology MESA+ Research
Institute University of Twente, The Netherlands}
\author{M. Elwenspoek}
\affiliation{Transducers Science and Technology MESA+ Research
Institute University of Twente, The Netherlands}

\date{\today}

\begin{abstract}
Voltage fluctuations of Platinum wires hosted by silicon nitride
beams were investigated. We considered four variants of the wires:
three with an adhesion layer and one with no adhesion layer. We
found that the presence of an adhesion layer changes the nature of
the power spectrum which is $1/f$ for wires with no adhesion
layers and $1/f^{3/2}$ for wires with an adhesion layer. We
attribute the value of the exponent $\alpha=3/2$ to the long range
diffusion of trapped hydrogen in the wires.
\end{abstract}

\pacs{Valid PACS appear here}
\maketitle
\begin{figure}
  \includegraphics[width=9 cm]{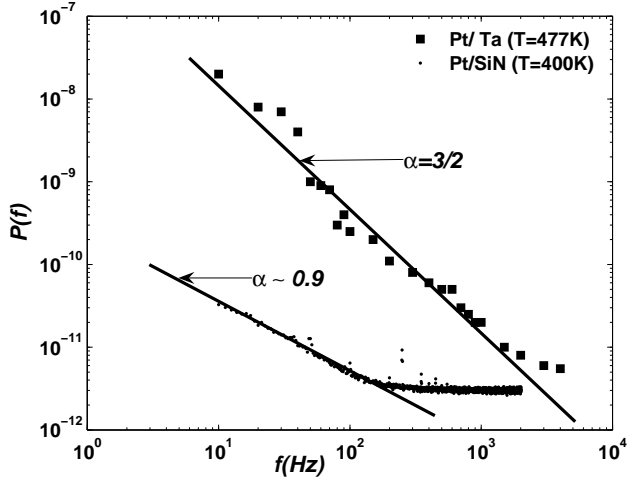}\\
  \caption{Voltage fluctuations power spectrum of Pt/Ta and Pt/SiN systems at T=477 K and T=400 K respectively.
  The value of $\alpha \simeq 3/2$ was common to all films with an adhesion layer. For clarity, these curves are shifted by few orders of magnitude.}\label{powerPt_Ta_SiN}
\end{figure}

\begin{figure}
  \includegraphics[width=9cm]{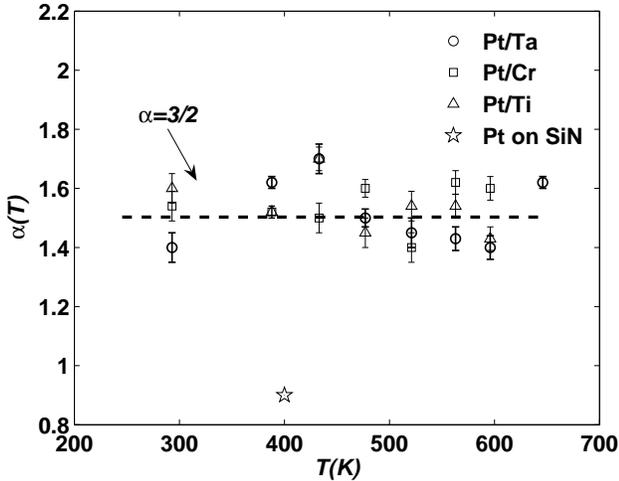}\\
  \caption{Plot of the exponent $\alpha$ as a function of the temperature for wires with and without an adhesion layer. Wires having an
  adhesion layer have a value of $\alpha$ consistent
with the value of 3/2 arising from the
  long range diffusions.}\label{Temp_alpha}
\end{figure}

 Low frequency electrical  noise in metallic films  has
been extensively investigated\cite{Dutta81,Scofield85,Weismann88}.
In a variety of metallic films,  the resulting noise spectra are
frequency dependent and are known as $1/f^\alpha$ spectrum with
the exponent $\alpha\simeq 1$. An important body of research into
the origin $1/f$ noise has emerged but a unifying picture is still
lacking. In the Dutta-Horn picture\cite{Dutta81}, $1/f$
 noise arises from equilibrium defect fluctuations
associated with the defects's motion which is thermally activated
and non-diffusive. These fluctuations have a characteristic time
scale $t_c$ which is itself sampled from a distribution function.
Other spectra of the form $1/f^{\alpha}$ were $\alpha \neq 1$ were
also found. In particular the value of $\alpha=3/2$ was attributed
to transport noise such as long range diffusion in metallic
films\cite{Scofield85_b,Nevins90,Scofield83} with hydrogen
impurities. This form of the spectrum  was also found in silver
films subject to electromigration damage\cite{Kar02}, where it was
argued that $1/f^{3/2}$ spectrum is attributed to the long-range
diffusion of
atoms through pathways opened during  electromigration damage. \\
The present work is concerned with the investigation of the noise
spectrum of a velocity sensor\cite{Bree}. The sensor consists of
two freely suspended silicon nitride beams on the top of which a
double layer of platinum  and an adhesive metallic layer (adhesion
layer) are deposited by sputtering. We show that  the presence of
the adhesion layer changes the nature of the power spectrum of the
voltage fluctuations of the wires. Without the adhesion layer, the
low frequency noise follows a power law with the usual power
spectrum $1/f^\alpha$, with $\alpha\simeq1$.  The presence of the
adhesion layer, changes this power spectrum to $1/f^\alpha$ with
$\alpha \simeq 3/2$. We attribute this effect to the long range
diffusion of hydrogen trapped in the Pt/adhesion layer structure.\\
To fabricate the velocity sensor, we consider four variants: three
variants were made by depositing  a 10 nm adhesion layer of
chromium (Cr), titanium (Ti) and tantalum(Ta) respectively
followed by a deposition of 150 nm of platinum. These variants
will be refereed to as Pt/Cr, Pt/Ti and Pt/Ta. The fourth variant
consist of a device without an adhesion layer thus a layer of 150
nm deposited directly on silicon nitride. This variant will be
refereed to as Pt/SiN. All films were deposited in a DC magnetron
sputtering system. The system has a rotating substrate holder,
with a distance of about 18 cm between substrate and target. After
the deposition of the adhesion layers, the samples stayed in vacuo
prior to the deposition of the platinum film. After patterning and
etching, the resistors were released with dimensions of 1 mm in
length and 4 $\mu m$ in width. Prior to experiments, the sensors
were annealed at $500^o$C. They consisted of two wires which are
200 $\mu m$ apart. The resistance of the wires was 300 $\Omega$ at
room temperature. As a reference in the experiments, a common
metal film resistor of 300 $\Omega$ was used. The output signal
was amplified twice by two amplifiers with an amplification factor
of 50 dB . The signals were recorded by a 20 bits AD-converter
with an input impedance of 30 K$\Omega$, and a flat frequency
response in the bandwidth (30 Hz, 20 kHz) . The recorded signal
was digitized and analyzed in a personal computer. The temperature
of the wires was varied from 293 K to 646 K, and the spectrum of
the detected output signal (in V$^2$/Hz) at each temperature was
computed. In figure \ref{powerPt_Ta_SiN} we show the frequency
dependence of the power spectrum of the voltage fluctuations for
Pt/SiN and Pt/Ta at temperatures 400 K and 477 K respectively. The
power spectrum scales with frequency as $S_V(f) \sim 1/f^\alpha$
for frequencies below 3kHz for Pt/Ta and below 200 Hz for Pt/SiN.
The values of $\alpha$ are  $\alpha \simeq 0.90$ and $ \alpha
\simeq 1.62$ for Pt/SiN and Pt/Ta respectively. These values were
determined by a regression fit on a double logarithmic plot of the
power spectrum versus frequency. The flat portions of the spectrum
correspond to white noise level. The value of $\alpha \simeq 1$
was also observed by Scofield and Mantese \cite{Scofield85} for
142 nm-thick platinum on sapphire at 400$^o$C.
For all wires (Pt/Ta, Pt/Ti, Pt/Cr and Pt/SiN) the power spectrum
follows the power law scaling $S_v(f) \sim 1/f^\alpha$. For wires
with an adhesion layer, the scaling was observed over a range of
two orders of magnitude in frequency and the value of the exponent
$\alpha$ was found to be close to 3/2 as shown in figure
\ref{Temp_alpha}. This exponent We found to be
temperature-independent (within the error range) over the
considered temperature range.

\begin{figure}
  \includegraphics[width=9cm]{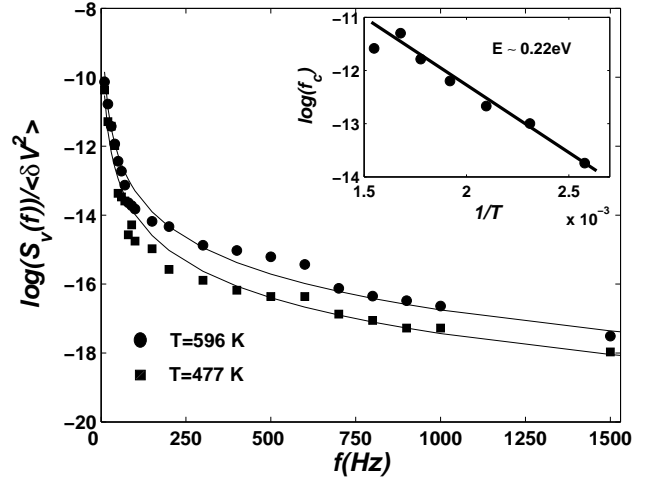}\\
  \caption{plot of $log(S_V(f))/<\delta V^2>$ versus f for Pt/Ta. The solid lines are the fit to the theoretical power spectrum. the inset shows
  $log(f_c)$ versus $1/T$}\label{powerfit}
\end{figure}

\begin{figure}
  \includegraphics[width=8cm]{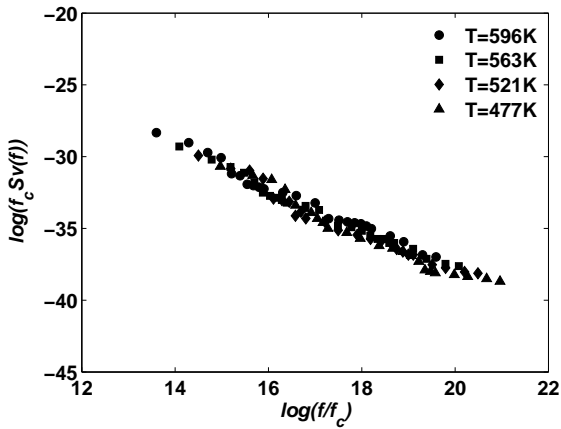}\\
  \caption{plots of $log(f_cS_V(f,T))$ versus $log(f/f_c(T))$ for Pt/Ta wire at different temperatures showing the collapse of the data
  on a single curve as predicted by the diffusion model.}\label{data_collapse}
\end{figure}

\begin{figure}
  \includegraphics[width=9cm]{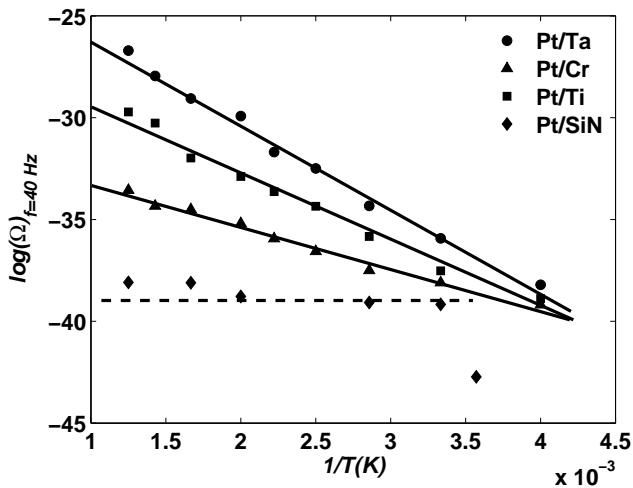}\\
  \caption{Plots of $\Omega(T)=f^\alpha S_V(f)$ vs $1/T$ showing a thermally activated process in films with an adhesion layer.}\label{gamma_vs_T}
\end{figure}

 The values of
$\alpha$ found in all wires with an adhesion layer are consistent
with the value found for resistance fluctuations in  Niobium
films\cite{Scofield85_b}
 $\alpha =3/2$. Scofield and Webb\cite{Scofield85_b} proposed that hydrogen diffusion in Niobium films modulates the
 resistance of the films where fluctuations of the number of hydrogen ions in the film segment induce the
 fluctuations of the resistance, i.e. $\delta r\propto \delta N$, where the fluctuation $\delta N$ obeys the diffusion equation in the
 fixed volume of the film's segment. In the framework of this model\cite{Scofield85_b,Voss76},
 a diffusing species in the film of length $L$, will generate a power spectrum of the form $fS_V(f)/<\delta V^2>=2P(x)-P(2x)$, where
  $x=(f/f_c)^{1/2}$, $f_c=D_0\exp(-E/K_BT)/\pi L^2$ and
\begin{equation}\label{eq1}
  P(x)=\left(1-e^{-x}\left(sinx+cosx\right)\right) /(\pi x)
  \end{equation}
  here,  $<\delta V^2>$ is the voltage variance.
 We first assume that the fluctuations in the wires with the adhesion layer, are described by the one dimensional diffusion process mentioned above
 and focus on  the Pt/Ta wires. To determine the characteristic frequency $f_c$ for Pt/Ta, we perform
a nonlinear regression fit of the data to the theoretical
expression of the one dimensional power spectrum (\ref{eq1}) in
the working frequency range of the sensor. In figure
\ref{powerfit} we show the plot of $log(S_V(f))/(<\delta V^2>)$
versus $f$ as well as the
 theoretical fit to the data for two different temperatures $T=596K$ and $T=477K$. We have determined
 the characteristic frequency at various temperatures: 596K, 563K,
 521K, 477K and 433K. The inset of figure \ref{powerfit} shows
 the plot of the logarithm of $f_c$ versus $1/T$, the slope gives
 the activation energy and the intercept gives the diffusion coefficient preexponential factor for
 Pt/Ta,
 which are $E_{Pt/Ta} \simeq 0.22 \pm 0.04$ and $D_0 \simeq 2.4 \times
 10^{-5} cm^2s^{-1}$ for $L=1mm$.  The value found for the activation energy is very
 close to the value derived directly from the semi-logarithmic plot of
 $S_V(40Hz)$ (figure \ref{gamma_vs_T}) as a function of $1/T$, which is related to
 $E_{Pt/Ta}$, by the relation $E'_{Pt/Ta}=(3/2)E_{Pt/Ta}$. We found
 $E'_{Pt/Ta}=0.35 \pm 0.02$ eV which gives $E_{Pt/Ta} \simeq 0.23$ eV. The
 evidence of the diffusion process can be shown by plotting
 $log(f_cS_V(f,T))$ versus $log(f/f_c(T))$ (figure \ref{data_collapse}) at different
 temperatures for Pt/Ta wires, which shows a data collapse into a single curve, as expected from a diffusion process.


In figure \ref{gamma_vs_T} we plot the logarithm of the quantity
$\Omega(T)=f^\alpha S_V(f)$ as a function of $1/T$ at $40 Hz$. The
plots corresponding to wires with an adhesion layer are
reminiscent of a thermally activated process. The activation
energies found for Pt/Ti and Pt/Cr wires are $0.190 \pm 0.02$ eV
and $0.118 \pm 0.01$ eV respectively. No evidence of a thermally
activated process was found in wires with no adhesion layer as
clearly noticed in figure \ref{gamma_vs_T}.
 The value of the activation energy found for the Pt/Ta wires is very close to the value found for trapped hydrogen in Niobium films ($\simeq 0.23$ eV).
 Hydrogen can be trapped in metals by  a variety of defects such as voids, high strain field around dislocations, grain boundaries or impurities\cite{Wert78}.
 Evanescent trapping of hydrogen by these defects increases the the activation energy to $E=E_0+E_b$, where $E_0$ is
the activation energy in the absence of traps and $E_b$ is the
energy associated with traps. Thin films such as those studied
here have different structure than bulk metals.
 In Pt/Ti films, ambient oxygen and titanium can diffuse through platinum films where they undergo chemical reactions to form TiO$_x$\cite{KO92,Olowolafe93} or
 PtTi$_x$ components. It was also reported that for thin titanium layers, the titanium
 oxide phase is located at the boundaries of platinum
 grains\cite{KO92}.
Tantalum hardly diffuses into the Pt layer, and consequently the
oxidation of Ta occurs mainly at the interface
Pt/Ta\cite{Maeder98}. As a result,
 stress induced defects are less present in Pt/Ta in comparison with Pt/Ti.
 Whereas chromium also has these effects (as for Pt/Ti), it also has a strong tendency to react with Pt to form an eutecticum,
 even at rather low temperatures\cite{Terblanche94}.
We believe that the observed diffusion possess originates from the
trapped hydrogen in the wires and that different activation
energies are associated with the difference in the nature of
scattering defects in each wire. The trapped hydrogen in the wires
might originate from sputtering or processing of  the sensors. For
example, it is well-known that significant amounts of hydrogen are
produced during KOH-etching; this is used to release the
SiN-beams. The absence of the long range diffusion in the Pt
without an adhesion layer suggests that hydrogen does not undergo
long range diffusion in this wire. In fact, hydrogen effusion is
very likely during high temperature annealing, which is
facilitated by the porous nature of Pt subject to high
temperatures. Our experiments show that there is less excess noise
in Pt without an adhesion layer than in Pt with an adhesion layer.
  The noise in Pt wires is 1/f and no evidence of thermal activation of this process was observed.
This can not be explained by the "defect motion" model of
Dutta-Horn.\\ The authors would like to thank R. Tiggelaar for
useful discussions.


\end{document}